\documentclass[12pt]{article}
\usepackage{amssymb,amsmath,epsfig}

\begin{document}

\title{\bf Dust Static Spherically Symmetric Solution in $f(R)$ Gravity}

\author{Muhammad Sharif \thanks{E-mail: msharif@math.pu.edu.pk} and Hafiza Rizwana
Kausar
\thanks{E-mail: rizwa\_math@yahoo.com}\\
Department of Mathematics, University of the Punjab,\\
Quaid-e-Azam Campus, Lahore-54590, Pakistan.}

\date{}
\maketitle

\begin{abstract}
In this paper, we take dust matter and investigate static
spherically symmetric solution of the field equations in metric
$f(R)$ gravity. The solution is found with constant Ricci scalar
curvature and its energy distribution is evaluated by using
Landau-Lifshitz energy-momentum complex. We also discuss the
stability condition and constant scalar curvature condition for
some specific popular choices of $f(R)$ models in addition to
their energy distribution.
\end{abstract}


\section{Introduction}

Our latest data from different sources, such as Cosmic Microwave
Background Radiations (CMBR) and supernova survey indicates that
energy composition of the universe is the following: $4\%$
ordinary matter, $20\%$ dark matter and $76\%$ dark energy
$^{1)}$. The dark energy has large negative pressure, while the
pressure of the dark matter is negligible. The current accelerated
expansion of the universe may be due to the presence of dark
energy so-called effective cosmological constant. There are
various directions aimed to construct the acceptable dark energy
model. For example, quintessence or phantom models, dark fluid
with complicated equation of state, String or M-theory, higher
dimensions, more complicated field theories, etc. Despite the
number of attempts, there is still no satisfactory explanation
about the origin of dark energy.

The $f(R)$ theory of gravity provides the very natural
gravitational alternative for dark energy. The cosmic acceleration
can be directly explained by taking any negative power of the
curvature (like $\frac{1}{R}$ term) $^{2)}$. In this way, this
theory helps in modification of the model to achieve the
consistency with the experimental tests of solar system. A
unification of the early time inflation and late time acceleration
$^{3)}$ is allowed in $f(R)$ theory. It is also very useful in
high energy physics for explaining the hierarchy problem and
unification of GUTs with gravity $^{4)}$.

This theory has produced a great number of papers in recent years,
for example, $^{5-7)}$. Several features including solar system
test $^{8)}$, Newtonian limit $^{9)}$, gravitational stability
$^{10)}$ and singularity problems $^{11)}$ are exhaustively
discussed. Much work has been devoted to place constraints on
$f(R)$ models using the observation of CMB anisotropies and galaxy
power spectrum $^{12)}$. Kobayashi and Maeda $^{13)}$ have studied
relativistic stars in this theory gravity. It is shown explicitly
that stars with strong gravitational fields develop curvature
singularity and hence are prohibited. Erickcek \textit{et al.}
$^{14)}$ found the unique exterior solution for a stellar object
by matching it with interior solution in the presence of matter
sources. Kainulainen \textit{et al.} $^{15)}$ studied the interior
spacetime of stars in Palatini $f(R)$ gravity.

The vacuum solutions of the field equations in metric $f(R)$
gravity has attracted many people. Since the spherically symmetry
plays a fundamental role in understanding the nature of gravity,
most of the solutions are discussed in this context.
Multam$\ddot{a}$ki and Vilja $^{16)}$ investigated static
spherically symmetric vacuum solutions of the field equations. It
is shown that solution with constant scalar curvature corresponds
to Schwarzschild de Sitter spacetime for a specific choice of
constants of integration. Caram$\hat{e}$s and Bezerra $^{17)}$
discussed spherically symmetric vacuum solutions in higher
dimensions. Capozziello \emph{et al.} $^{18)}$ analyzed
spherically symmetric solution using Noether symmetry.

Azadi \textit{et al.} $^{19)}$ have studied cylindrically
symmetric solutions in Weyl coordinates. They have shown that
constant curvature solutions reduce to only one member of the Tian
family in General Relativity (GR). Momeni $^{20)}$ has found that
the exact constant scalar curvature solution in cylindrical
symmetry is applicable to the exterior of a string. In a recent
work, Sharif and Shamir $^{21)}$ have studied exact solutions of
Bianchi types $I$ and $V$ spacetimes in $f(R)$ theory of gravity.
Multam$\ddot{a}$ki and Vilja $^{22)}$ investigated non-vacuum
solutions by taking perfect fluid. It is found that for a given
matter distribution and equation of state, one cannot determine
the function form of $f(R)$. Hollestein and Lobo $^{23)}$ explored
exact solutions of the field equations coupled to non-linear
electrodynamics.

The energy localization in GR is a serious and long standing issue
but without any definite answer. The well-known energy-momentum
prescriptions are given by Landau-Lifshitz, Einstein, Bergmann,
Papapetrou, Goldberg and M$\o$ller. Virbhadra \textit{et al.}
$^{24)}$ showed that five different energy-momentum complexes
yield the same energy distribution for any Kerr-Schild class
metric. Chang \textit{et al.} proved that every energy-momentum
complex is associated with a Hamiltonian boundary term $^{25)}$.
Recently, this problem has been attempted in alternative theories
of gravity. In this connection, reasonable amount of work has been
published $^{26)}$ in teleparallel theory of gravity.
Multam$\ddot{a}$ki \textit{et al.} $^{27)}$ are the pioneers to
discuss energy problem in $f(R)$ theory of gravity. They presented
the concept of energy-momentum complex (EMC) in this theory and
generalized the Landau-Lifshitz energy-momentum complex. The
prescription is used to evaluate energy-momentum for the
Schwarzschild-de-Sitter spacetime. Recently, Sharif and Shamir
$^{28)}$ found energy densities of plane symmetric and cosmic
string spacetimes. They also discussed the stability conditions.

The purpose of this paper is two fold: Firstly, we study non-vacuum
static spherically symmetric solutions of the field equations using
metric $f(R)$ gravity in the presence of dust fluid. Secondly, we
use generalized Landau-Lifshitz energy-momentum complex to evaluate
energy density for constant scalar curvature solution. The paper is
organized as follows. In section \textbf{2}, we present spherically
symmetric field equations and some of the relevant quantities.
Section \textbf{3} is devoted to study the non-trivial solution of
the field equations. In section \textbf{4}, we calculate the
generalized Landau-Lifshitz energy-momentum complex for constant
scalar curvature solution and also discuss some well-known $f(R)$
models in this context. In the last section \textbf{5}, we summarize
and discuss the results.

\section{Field Equations in $f(R)$ Gravity}

The action in $f(R)$ gravity is given by $^{29)}$
\begin{equation}\label{2.1}
S=\int d^{4}x\sqrt{-g}\left[\frac{f(R)}{2\kappa}+L_{M}\right],
\end{equation}
where $f(R)$ is a function of the Ricci scalar and $L_{M}$ is the
matter Lagrangian depending upon the metric $g_{\mu\nu}$ and the
matter fields. Variation of this action with respect to the metric
tensor leads to the following  fourth order partial differential
equations
\begin{equation}\label{2.2}
F(R) R_{\mu\nu} - \frac{1}{2}f(R)g_{\mu\nu}-\nabla_{\mu}
\nabla_{\nu}F(R)+ g_{\mu\nu} \Box F(R)= \kappa T_{\mu\nu},
\end{equation}
where $F(R)\equiv df(R)/dR,~\Box \equiv \nabla^{\mu}\nabla_{\mu}$
with $\nabla_{\mu}$ representing the covariant derivative and
$\kappa(=8\pi)$ is the coupling constant in gravitational units.
Taking trace of the above equation, we obtain
\begin{equation}\label{2.3}
F(R) R - 2f(R)+ 3\Box F(R)= 8\pi T.
\end{equation}
Here $R$ and $T$ are related differentially and not algebraically as
in GR ($R=-\kappa T$). This indicates that the field equations
of $f(R)$ gravity will admit a larger variety of solutions than does
GR. Further, $T=0$ does no longer implies $R=0$ in this theory.

The Ricci scalar curvature function $f(R)$ can be expressed in terms
of its derivative as follows
\begin{equation}\label{2.4}
f(R)=\frac{-8\pi T+ F(R) R+ 3\Box F(R)}{2}.
\end{equation}
This is used to study various aspects of $f(R)$ gravity,
particularly its stability, weak field limit etc. Substituting this
value of $f(R)$ in Eq.(\ref{2.2}), we obtain
\begin{equation}\label{2.5}
\frac{F(R) R-\Box F(R)-8\pi T}{4}=\frac{F(R)
R_{\mu\mu}-\nabla_{\mu} \nabla_{\mu}F(R)-8\pi T_{\mu
\mu}}{g_{\mu\mu}}.
\end{equation}
In the above equation, the expression on the left hand side is
independent of the index $\mu$, so the field equations can be
written as
\begin{equation}\label{2.6}
A_{\mu}=\frac{F(R) R_{\mu\mu}-\nabla_{\mu} \nabla_{\mu}F(R)-8\pi
T_{\mu \mu}}{g_{\mu\mu}}.
\end{equation}
Notice that $A_{\mu}$ is not a 4-vector rather just a notation for
the traced quantity.

\subsection{Spherically symmetric spacetime}

We take the following static spherically symmetric spacetime
\begin{equation}\label{3.1}
ds^2=A(r)dt^2-B(r)dr^2-r^2(d\theta^2+\sin^2\theta d\phi^2).
\end{equation}
The components of the Ricci tensor are
\begin{eqnarray}
R_{00}&=&-\frac{1}{4B}[-2A''+\frac{A'^2}{A}+\frac{A'B'}{B}- 4\frac{A'}{r}],\\
R_{11}&=&\frac{1}{4A}[-2A''+\frac{A'^2}{A}+\frac{A'B'}{B}+ 4\frac{B'A}{Br}],\\
R_{22}&=&-\frac{1}{2}[\frac{A'r}{AB}-\frac{B'r}{B^2}-2+
\frac{2}{B}],\\
R_{33}&=&{\sin}^2{\theta}R_{22}.
\end{eqnarray}
The corresponding Ricci scalar is
\begin{equation}\label{3.3}
R=\frac{-2}{r^2B}[1-B+\frac{r^2A''}{2A}+\frac{A'}{A}(r-\frac{r^2A'}{4A})
-\frac{B'}{B}(r+\frac{r^2A'}{4A})],
\end{equation}
where prime denotes derivative with respect to the radial
coordinate $r$. The dust energy-momentum tensor is given as
\begin{equation}\label{3.4}
T_{\mu\nu}=\rho u_\mu u_\nu,
\end{equation}
where $u_\mu=\delta^0_\mu$ is the four-velocity in co-moving
coordinates and $\rho$ is the density. Since Eq.(\ref{2.5}) is
independent of index $\mu$, so $A_{\mu}-A_{\nu}=0$ for all
$\mu,~\nu$ and yields the following two independent equations
\begin{eqnarray} \label{3.5}
&&-\frac{F''}{B}+\frac{F'}{2B}[\frac{A'}{A}+\frac{B'}{B}]+F[\frac{A'}{B
A r} +\frac{B'}{B^2r}] -\frac{8\pi\rho}{A}=0,\\\label{3.6}
&&F'[\frac{A'}{2BA}-\frac{1}{B
r}]+F[\frac{A''}{2AB}-\frac{A'^2}{4A^2B}
-\frac{A'B'}{4AB^2}+\frac{A'}{2ABr} +\frac{B'}{2B^2r}\nonumber\\&&
+\frac{1}{r^2}-\frac{1}{Br^2}]-\frac{8\pi\rho}{A}=0.
\end{eqnarray}
Thus we get a system of two non-linear differential equations with
four unknown functions, namely, $F(r),~\rho(r),~A(r)$ and $B(r)$.

\section{Solution of the Field Equations}

This section is devoted to discuss solution of the field equations
by assuming constant scalar curvature which are directly involved
in explaining the accelerating universe. In order to take into
account acceleration of the present universe, we need to take very
small value of the constant to $f(R)$ $^{3)}$.

The conservation law of energy-momentum tensor,
$T^{\nu}_{\mu;\nu}=0$, for dust matter $^{22)}$ gives
$A=\textmd{constant}=A_0~(\textmd{say})$. Thus the system of field
equations (\ref{3.5}) and (\ref{3.6}) is reduced to three unknowns
with the following two non-linear differential equations
\begin{eqnarray} \label{3.7}
&&\frac{-1}{B}F''+\frac{B'}{2B^2}F'+\frac{B'}{B^2r}F
-\frac{8\pi\rho}{A_o}=0,\\\label{3.8} &&\frac{-1}{B
r}F'+[\frac{B'}{2B^2r}+\frac{1}{r^2}-\frac{1}{Br^2}]F-\frac{8\pi\rho}{A_0}=0.
\end{eqnarray}
Using the assumption of constant scalar curvature ($R=R_{0}$), i.e.,
$F(R_0)=\textmd{constant}$, the field equations become
\begin{eqnarray} \label{4.1}
&&\frac{B'}{B^2r}F(R_0)-\frac{8\pi\rho}{A_0}=0,\\\label{4.2}
&&(\frac{B'}{2B^2r}
+\frac{1}{r^2}-\frac{1}{Br^2})F(R_0)-\frac{8\pi\rho}{A_0}=0.
\end{eqnarray}
Now we have two differential equations with two unknowns, $B(r)$ and
$\rho(r)$. Subtracting Eq.(\ref{4.2}) from Eq.(\ref{4.1}), we have
an ordinary differential equation in terms of $B(r)$ as follows
\begin{equation}\label{4.3}
B'r+2B-2B^2=0
\end{equation}
which has the following solution
\begin{equation}\label{4.4}
B(r)=\frac{1}{1-c_1r^2},
\end{equation}
where $c_1$ is a constant. Inserting this value of $B$ in any of the
above equations, it follows that
\begin{equation}\label{4.5}
\rho={\frac{c_1A_0 F(R_0)}{4\pi}}=\rho_0
\end{equation}
which is purely a constant. We would like to mention here that if
matter density and scalar curvature are constant and also $\rho_0$
depends on gravitational constant and effective cosmological
constant $^{30)}$, then the field equations become equivalent to
the Einstein field equations. The spacetime for constant curvature
solution takes the following form
\begin{equation}\label{4.6a}
ds^2=A_0dt^2-\frac{1}{1-c_1r^2}dr^2-r^2(d\theta^2+\sin^2\theta
d\phi^2).
\end{equation}
This solution corresponds to the well-known
Tolman-Oppenheimer-Volkoff (TOV) spacetime when density is
constant and pressure is neglected $^{31)}$. For all ordinary,
non-relativistic stars where $p<<\rho$ (for example at the center
of the sun), we can neglect pressure and can consider an idealized
object with a constant density.

The scalar curvature turns out to be $R_0=6 c_1$ and hence
Eq.(\ref{2.3}) yields
\begin{equation}\label{4.6b}
f(R_0)=\frac{1}{2}(\frac{-8\pi\rho_0}{A_0}+F(R_0)R_0).
\end{equation}
Substituting the values of $\rho_0$ and $R_0$, it follows that
\begin{equation}\label{4.6c}
f(R_0)=2c_1f'(R_0).
\end{equation}
For the acceptability of any $f(R)$ model, it is necessary to
satisfy this equation. In the next section, we check this condition
for some well-known $f(R)$ models and then calculate energy density
using the generalized Landau-Lifshitz energy-momentum complex.

In view of the above information, the universe could start from
inflation driven by the effective cosmological constant at the early
stage where curvature is very large. With the passage of time, the
effective cosmological constant also becomes smaller corresponding
to the smaller curvature. After that time, the density of matter or
radiations become small and curvature goes to constant value $R_0$.
Thus expansion could start and cosmological constant can be
identified as $f(R_0)$ in the present accelerating era.

\section{Landau-Lifshitz Energy-Momentum Complex}

Now we evaluate energy density for the constant scalar curvature
solution. For this purpose, we use the generalized Landau-Lifshitz
energy-momentum complex. We would like to mention here that this
energy-momentum complex is valid only for those solutions which have
constant scalar curvature.

The generalized Landau-Lifshitz energy-momentum complex $^{27)}$
is given as follows
\begin{equation}\label{A1}
\tau^{\mu\nu}=f'(R_0)\tau^{\mu\nu}_{LL}+\frac{1}{6\kappa}
(f'(R_0)R_0-f(R_0))\frac{\partial}{\partial
x^\lambda}(g^{\mu\nu}x^\lambda - g^{\mu\lambda}x^\nu).
\end{equation}
Here $\tau^{\mu\nu}_{LL}$ is the Landau-Lifshitz energy-momentum
complex evaluated in the framework of GR and is
given as
\begin{equation}\label{A2}
\tau^{\mu\nu}_{LL}=(-g)(T^{\mu\nu}+t^{\mu\nu}_{LL}),
\end{equation}
where $t^{\mu\nu}_{LL}$ can be obtained by the following formula
\begin{eqnarray}\nonumber
t^{\mu\nu}_{LL}&=&\frac{1}{2\kappa}[(2\Gamma^\gamma_{\alpha\beta}\Gamma^\delta_{\gamma\delta}
-\Gamma^\gamma_{\alpha\delta}\Gamma^\delta_{\beta\gamma}
-\Gamma^\gamma_{\alpha\gamma}\Gamma^\delta_{\beta\delta})(
g^{\mu\alpha}g^{\nu\beta}-g^{\mu\nu}g^{\alpha\beta})\\\nonumber
&+&g^{\mu\alpha}g^{\beta\gamma}(\Gamma^\nu_{\alpha\delta}\Gamma^\delta_{\beta\gamma}
+\Gamma^\nu_{\beta\gamma}\Gamma^\delta_{\alpha\delta}
-\Gamma^\nu_{\gamma\delta}\Gamma^\delta_{\alpha\beta}
-\Gamma^\nu_{\alpha\beta}\Gamma^\delta_{\gamma\delta})\\\nonumber
&+&g^{\nu\alpha}g^{\beta\gamma}(\Gamma^\mu_{\alpha\delta}\Gamma^\delta_{\beta\gamma}
+\Gamma^\mu_{\beta\gamma}\Gamma^\delta_{\alpha\delta}
-\Gamma^\mu_{\gamma\delta}\Gamma^\delta_{\alpha\beta}
-\Gamma^\mu_{\alpha\beta}\Gamma^\delta_{\gamma\delta})\\\label{A3}
&+&g^{\alpha\beta}g^{\gamma\delta}(\Gamma^\mu_{\alpha\gamma}\Gamma^\nu_{\beta\delta}
-\Gamma^\mu_{\alpha\beta}\Gamma^\nu_{\gamma\delta})].
\end{eqnarray}
The $00$-component of Eq.(\ref{A1}) yields
\begin{eqnarray}\nonumber
\tau^{00}&=&f'(R_0)\tau^{00}_{LL}+\frac{1}{6\kappa}
(f'(R_0)R_0-f(R_0))\frac{\partial}{\partial
x^\lambda}(g^{00}x^\lambda - g^{0\lambda}x^0)\\\label{A4}
&=&f'(R_0)\tau^{00}_{LL}+\frac{1}{6\kappa}
(f'(R_0)R_0-f(R_0))(\frac{\partial g^{00}}{\partial
x^\lambda}x^\lambda+3 g^{00})
\end{eqnarray}
with
\begin{equation}\label{A5}
\tau^{00}_{LL}=(-g)(T^{00}+t^{00}_{LL}).
\end{equation}
We can find $t^{00}_{LL}$ from Eq.(\ref{A3}) as follows
\begin{eqnarray*}\nonumber
t^{00}_{LL}&=&\frac{1}{2\kappa}[(2\Gamma^\gamma_{\alpha\beta}\Gamma^\delta_{\gamma\delta}
-\Gamma^\gamma_{\alpha\delta}\Gamma^\delta_{\beta\gamma}
-\Gamma^\gamma_{\alpha\gamma}\Gamma^\delta_{\beta\delta})(
g^{0\alpha}g^{0\beta}-g^{00}g^{\alpha\beta})\\\nonumber
&+&g^{0\alpha}g^{\beta\gamma}(\Gamma^0_{\alpha\delta}\Gamma^\delta_{\beta\gamma}
+\Gamma^0_{\beta\gamma}\Gamma^\delta_{\alpha\delta}
-\Gamma^0_{\gamma\delta}\Gamma^\delta_{\alpha\beta}
-\Gamma^0_{\alpha\beta}\Gamma^\delta_{\gamma\delta})\\\nonumber
&+&g^{0\alpha}g^{\beta\gamma}(\Gamma^0_{\alpha\delta}\Gamma^\delta_{\beta\gamma}
+\Gamma^0_{\beta\gamma}\Gamma^\delta_{\alpha\delta}
-\Gamma^0_{\gamma\delta}\Gamma^\delta_{\alpha\beta}
-\Gamma^0_{\alpha\beta}\Gamma^\delta_{\gamma\delta})\\
&+&g^{\alpha\beta}g^{\gamma\delta}(\Gamma^0_{\alpha\gamma}\Gamma^0_{\beta\delta}
-\Gamma^0_{\alpha\beta}\Gamma^0_{\gamma\delta})],\quad\alpha,~\beta,~\gamma,~\delta=0,1,2,3.
\end{eqnarray*}
which finally gives
\begin{eqnarray}\nonumber
t^{00}_{LL}&=&\frac{1}{16\pi
A_0}[(1-c_1r^2)\{\frac{2c_1}{1-c_1r^2}+\frac{rc_1}{1-c_1r^2}
\cot\theta-2r^2+\frac{1}{r}\cot\theta\}\\\nonumber
&+&\frac{1}{r^2}\{\frac{2c_1}{1-c_1r^2}-\frac{1}{r}
\cot\theta-2(1-c_1r^2)-2\cot^2\theta\}\\\nonumber
&+&\frac{1}{r^2\sin^2\theta}
\{\frac{2c_1}{1-c_1r^2}+\frac{4}{r^2}+\frac{6}{r}\cot\theta+\frac{2c_1}{1-c_1r^2}
\cot\theta\\\label{A7}
&+&2\cot^2\theta-2\cos^2\theta\}].
\end{eqnarray}
Inserting this value in Eq.(\ref{A5}) and then substituting the
resulting equation in Eq.(\ref{A4}), it follows that
\begin{eqnarray}\nonumber
\tau^{00}&=&f'(R_0)\frac{A_0r^4\sin^2\theta
}{1-c_1r^2}[\frac{\rho}{A_0^2}+\frac{1}{16\pi
A_0}\{(1-c_1r^2)\{\frac{2c_1}{1-c_1r^2}\\\nonumber
&+&\frac{rc_1}{1-c_1r^2}\cot\theta-2r^2+\frac{1}{r}\cot\theta\}
+\frac{1}{r^2}\{\frac{2c_1}{1-c_1r^2}-\frac{1}{r}\cot\theta\\\nonumber
&-&2(1-c_1r^2)-2\cot^2\theta\}+\frac{1}{r^2\sin^2\theta}
\{\frac{2c_1}{1-c_1r^2}+\frac{4}{r^2}+\frac{6}{r}\cot\theta\\\nonumber
&+&\frac{2c_1}{1-c_1r^2}\cot\theta
+2\cot^2\theta-2\cos^2\theta\}\}]+\frac{1}{16\pi A_0}
(f'(R_0)R_0\\\label{A8}&-&f(R_0)).
\end{eqnarray}
This is the energy density satisfying the condition of constant
scalar curvature. We can evaluate this quantity for different
well-known $f(R)$ models. Also, we check the validity and the
stability condition for these models in the context of cosmology.

\subsection{Energy Density of the First Model}

First of all we evaluate energy density for the model
$f(R)=R+\epsilon R^2$, where $\epsilon$ is any positive real number.
The stability criteria for this model is restricted to $\epsilon<0$
which corresponds to $f^{''}(R)>0$. For $\epsilon=0$, GR is
recovered in which black holes are stable classically but not
quantum mechanically due to Hawking radiations. Since such features
also found in $f(R)$ gravity, hence the classical stability
condition for the Schwarzschild black hole can be enunciated as
$f^{''}(R)>0$. It is interesting to mention here that this model
satisfies the condition of constant scalar curvature for a specific
value of the constant $c_1$, i.e., $c_1=\frac{1}{3\epsilon}$.
Further, the stability condition for this model $^{32)}$,
$\frac{1}{\epsilon(1+2\epsilon R_0)}=\frac{1}{5\epsilon}>0$, is also
satisfied.

The $00$-component of the generalized EMC takes the form
\begin{eqnarray}\nonumber
\tau^{00}&=&(1+12\epsilon
c_1)\frac{A_0r^4\sin^2\theta}{1-c_1r^2}[\frac{\rho}{A_0^2}+\frac{1}{16\pi
A_0}\{(1-c_1r^2)\\\nonumber
&\times&\{\frac{2c_1}{1-c_1r^2}+\frac{rc_1}{1-c_1r^2}\cot\theta
-2r^2+\frac{1}{r}\cot\theta\}
+\frac{1}{r^2}\{\frac{2c_1}{1-c_1r^2}\\\nonumber &-&\frac{1}{r}
\cot\theta-2(1-c_1r^2)-2\cot^2\theta\} +\frac{1}{r^2\sin^2\theta}
\{\frac{2c_1}{1-c_1r^2}+\frac{4}{r^2}\\\nonumber
&+&\frac{6}{r}\cot\theta +\frac{2c_1}{1-c_1r^2}\cot\theta
+2\cot^2\theta-2\cos^2\theta\}\}]+\frac{9\epsilon c_1^2}{4\pi A_0}.
\end{eqnarray}

\subsection{Energy Density of the Second Model}

Here we use the model $f(R)=R-\frac{a}{R}-bR^2$ to evaluate energy
density, where $a$ and $b$ are any real numbers. This model involves
the negative power of the curvature which supports the cosmic
acceleration. In this way, any negative power of the curvature can
be taken into account to achieve the consistency with experimental
tests of Newtonian gravity. However, the model involving such term
might not satisfy the stability conditions which can be
significantly improved by adding square terms of the scalar
curvature. For $R=R_0=6c_1$, we have
\begin{equation}
f(R_0)=6c_1-\frac{a}{6c_1}-36bc_1^2,\quad
f'(R_0)=1+\frac{a}{36c_1^2}-12bc_1.
\end{equation}
The constant curvature condition implies that both the parameters
$a$ and $b$ are related by the following expression
\begin{equation}\label{m3}
1-\frac{a}{18c_1^2}-3bc_1=0.
\end{equation}
Further, the stability condition $^{27)}$ $f''(R_0)>0$ yields
$a+b(R_0)^3\geq0$ which is satisfied for $c_1^2\geq \frac{5a}{72}$.
Thus the model is acceptable for such choice of $c_1$.

Consequently, the energy density takes the form
\begin{eqnarray}\nonumber
\tau^{00}&=&(1+\frac{a}{36c_1^2}-12bc_1)\frac{A_0r^4\sin^2\theta
}{1-c_1r^2}[\frac{\rho}{A_0^2}+\frac{1}{16\pi
A_0}\{(1-c_1r^2)\\\nonumber&&
\{\frac{2c_1}{1-c_1r^2}+\frac{rc_1}{1-c_1r^2}\cot\theta
-2r^2+\frac{1}{r}\cot\theta\}\\\nonumber
&+&\frac{1}{r^2}\{\frac{2c_1}{1-c_1r^2}-\frac{1}{r}
\cot\theta-2(1-c_1r^2)-2\cot^2\theta\}\\\nonumber
&+&\frac{1}{r^2\sin^2\theta}
\{\frac{2c_1}{1-c_1r^2}+\frac{4}{r^2}+\frac{6}{r}\cot\theta
+\frac{2c_1}{1-c_1r^2}\cot\theta\\\label{A10}
&+&2\cot^2\theta-2\cos^2\theta\}\}]+\frac{1}{16\pi A_0}(\frac{a}
{3c_1}-36bc_1^2).
\end{eqnarray}

\subsection{Energy Density of the Third Model}

The model considered here is
$f(R)=R-(-1)^{n-1}\frac{a}{R^n}+(-1)^{m-1}bR^m$, where $m$ and $n$
are positive integers and $a,b$ are real numbers, and
is widely used in cosmology. A model of such type with
$m=1-\frac{\alpha}{2}$ with $\alpha$ depending upon the mass of the
galaxy is used to approximate galaxies by taking spherically
symmetric solutions. It is straightforward that one cannot fit the
data for all astronomical masses for a single choice of $f(R)$ as
$\alpha$ depends upon the mass of individual galaxy. For $R=R_0$,
the model becomes
\begin{eqnarray}
f(R)&=&\frac{(6c_1)^{n+1}-(-1)^{n-1}a+(-1)^{m-1}b(6c_1)^{m+n}}{(6c_1)^n},\\
f'(R_0)&=&\frac{(6c_1)^{n+1}+(-1)^{n-1}na+(-1)^{m-1}bm(6c_1)^{m+n}}{(6c_1)^{n+1}}.
\end{eqnarray}
For this form of $f(R)$, the constant curvature condition can be
written as
\begin{equation}
2(6c_1)^{n+1}-(-1)^{n-1}a(3+n)+(-1)^{m-1}b(3-m)(6c_1)^{m+n}=0.
\end{equation}
In particular, when $m=3$ or $b=0$, we get
\begin{equation}
a=\frac{(-1)^{n-1}2(6c_1)^{n+1}}{n+3},\quad n\neq3.
\end{equation}
It is worthwhile to mention here that the stability condition is
satisfied for this value of $a$.

The corresponding $00$-component gives
\begin{eqnarray}\nonumber
\tau^{00}&=&\frac{(6c_1)^{n+1}+(-1)^{n-1}na+(-1)^{m-1}bm(6c_1)^{m+n}}{(6c_1)^{n+1}}\\\nonumber
&\times&(\frac{A_0r^4\sin^2\theta
}{1-c_1r^2})[\frac{\rho}{A_0^2}+\frac{1}{16\pi
A_0}\{(1-c_1r^2)\\\nonumber
&\times&\{\frac{2c_1}{1-c_1r^2}+\frac{rc_1}{1-c_1r^2}\cot\theta
-2r^2+\frac{1}{r}\cot\theta\}\\\nonumber
&+&\frac{1}{r^2}\{\frac{2c_1}{1-c_1r^2}-\frac{1}{r}
\cot\theta-2(1-c_1r^2)-2\cot^2\theta\}\\\nonumber
&+&\frac{1}{r^2\sin^2\theta}
\{\frac{2c_1}{1-c_1r^2}+\frac{4}{r^2}+\frac{6}{r}\cot\theta
+\frac{2c_1}{1-c_1r^2}\cot\theta\\\nonumber
&+&2\cot^2\theta-2\cos^2\theta\}\}]+\frac{1}{16\pi
A_0}\\\label{A11}
&\times&\frac{(-1)^{n-1}a(n+1)+(-1)^{m-1}b(m-1)(6c_1)^{m+n}}{(6c_1)^n}.
\end{eqnarray}

\subsection{Energy Density of the Fourth Model}

This is an interesting model due to its logarithmic dependence on
curvature and it also satisfies the existence of relativistic stars.
It is given as follows, $f(R)=R-a\ln(\frac{|R|}{k})+(-1)^{m-1}bR^m$,
where $m$ is a positive integer, $k$ is a positive real number and
$a$ is any real number. For constant curvature $R_0$, we have
\begin{equation}
f(R)=6c_1-a\ln(\frac{6c_1}{k})+(-1)^{m-1}b(6c_1)^m
\end{equation}
and
\begin{equation}
f'(R_0)=\frac{6c_1-a+(-1)^{m-1}bm(6c_1)^m}{6c_1}.
\end{equation}
The constant curvature condition is expressed as
\begin{equation}
4c_1-a(\ln(\frac{6c_1}{k})-\frac{1}{3})+(-1)^{m-1}b(1-\frac{m}{3})(6c_1)^m=0.
\end{equation}
For $m=3$ or $b=0$, this reduces to
\begin{equation}
a=\frac{4c_1}{\ln(\frac{6c_1}{k})-\frac{1}{3}}.
\end{equation}\nonumber

The corresponding $00$-component is of the form
\begin{eqnarray}\nonumber
\tau^{00}&=&\frac{6c_1-a+(-1)^{m-1}b
m(6c_1)^{m-1}}{6c_1}\\\nonumber &\times&(\frac{A_0r^4\sin^2\theta
}{1-c_1r^2})[\frac{\rho}{A_0^2}+\frac{1}{16\pi
A_0}\{(1-c_1r^2)\\\nonumber
&\times&\{\frac{2c_1}{1-c_1r^2}+\frac{rc_1}{1-c_1r^2}\cot\theta
-2r^2+\frac{1}{r}\cot\theta\}\\\nonumber
&+&\frac{1}{r^2}\{\frac{2c_1}{1-c_1r^2}-\frac{1}{r}
\cot\theta-2(1-c_1r^2)-2\cot^2\theta\}\\\nonumber
&+&\frac{1}{r^2\sin^2\theta}
\{\frac{2c_1}{1-c_1r^2}+\frac{4}{r^2}+\frac{6}{r}\cot\theta
+\frac{2c_1}{1-c_1r^2}\cot\theta\\\nonumber
&+&2\cot^2\theta-2\cos^2\theta\}\}]+\frac{1}{16\pi
A_0}\\\label{A12}
&\times&\{a(\ln(\frac{6c_1}{k})-1)+(-1)^{m-1}b(m-1)(6c_1)^m\}.
\end{eqnarray}

\section{Outlook}

This paper investigates solution of static spherically symmetric
spacetime with non-trivial matter distribution. We have restricted
our analysis to the dust case and obtain solution with assumption of
constant scalar curvature. In addition, we have explored energy
localization problem using the generalized Landau-Lifshitz
energy-momentum complex in $f(R)$ gravity.

The scalar curvature for this solution turns out to be non-zero
constant. This leads to constant density of dust matter and
corresponds to the well-known Tolman-Oppenheimer-Volkoff spacetime
when density is constant and pressure is neglected $^{31)}$. Thus
it is interesting to discuss such solutions for idealized objects
where density is constant and pressure can be neglected.

For such solutions with constant curvature, we can use the
generalized Landau-Lifshitz energy-momentum complex to discuss
energy-momentum distribution. We have evaluated energy density for
this solution as well as for certain specific $f(R)$ models. It is
worthwhile to mention here that the well-known $f(R)$ models satisfy
the stability condition as well as constant scalar curvature
condition for this solution. The cosmological importance of all
these models is also discussed. The model with negative power of the
scalar curvature directly supports the cosmic acceleration. If we
choose the model with positive powers of the scalar curvature
(higher than 1), the mass of the scalar field can be adjusted to be
very large and the stability can be improved. For solutions with
zero scalar curvature, the generalized Landau-Lifshitz EMC coincides
with Landau-Lifshitz EMC in GR. This work adds some knowledge about
the longstanding and crucial problem of the localization of energy.
It gives the energy density expressions for important $f(R)$ models
which may help at some stage to overcome the theoretical
difficulties in the cosmological and astrophysical context.

The applicability of solutions could be tested by comparing with
constraints of the solar system and cosmology. It would be
interesting to investigate solutions for non-static spacetimes
with energy-momentum tensor of other types of fluids.

\vspace{0.25cm}

{\bf Acknowledgments}

\vspace{0.25cm}

We would like to thank the Higher Education Commission, Islamabad,
Pakistan for its financial support through the {\it Indigenous Ph.D.
5000 Fellowship Program Batch-III}. We are also grateful to the
Physical Society of Japan for financial support in publication.

\end{document}